\documentclass{PoS}
\usepackage{amsmath}

\newcommand{\Tr}{\mbox{Tr}}

\def\gtwid{{\,\raise.3ex\hbox{$>$\kern-.75em\lower1ex\hbox{$\sim$}}\,}}
\def\ltwid{{\,\raise.3ex\hbox{$<$\kern-.75em\lower1ex\hbox{$\sim$}}\,}}

\def\chpt{\raise0.4ex\hbox{$\chi$}PT}
\def\schpt{S\raise0.4ex\hbox{$\chi$}PT}
\def\rschpt{rS\raise0.4ex\hbox{$\chi$}PT}
\def\wchpt{W\raise0.4ex\hbox{$\chi$}PT}

\title{Low-lying Dirac operator eigenvalues, lattice effects and random matrix theory%
\footnote{PoS {\bf Lattice 2011} (2011) 103}}

\ShortTitle{Low-lying Dirac operator eigenvalues, lattice effects and random matrix theory}

\author{\speaker{Urs Heller}%
         \\
        American Physical Society, One Research Road, Ridge, NY 11961, USA\\
        E-mail: \email{heller@aps.org}}

\abstract{
Recently, random matrix theory predictions for the distribution of
low-lying Dirac operator eigenvalues have been extended to include lattice
effects for both staggered and Wilson fermions. We computed low-lying
eigenvalues for the Hermitian Wilson-Dirac operator and for improved
staggered fermions on several quenched ensembles with size $\approx 1.5$
fm. Comparisons to the expectations from RMT with lattice effects included
are made. Wilson RMT describes our Wilson data nicely. For improved
staggered fermions we find strong indications that taste breaking effects
on the low-lying spectrum disappear in the continuum limit, as expected
from staggered RMT.
}

\FullConference{XXIX International Symposium on Lattice Field Theory \\
		 July 10-16 2011\\
		 Squaw Valley, Lake Tahoe, California}

\begin{document}

\section{Chiral perturbation theory including lattice artifacts}

In recent years, chiral perturbation theory (\chpt) has been extended to
include lattice artifacts for both staggered \cite{SCHiPT} and Wilson
fermions \cite{WCHiPT}. In this talk, we will be concerned only with
the $\epsilon$-regime of \chpt\ where the zero momentum modes dominate.
Thus, we consider only the zero momentum part of the chiral Lagrangian
\begin{equation}
{\cal L} =  -\frac{1}{2} m \Sigma \Tr \left( U + U^\dagger \right)
 + a^2 {\cal V} ~.
\label{eq:L_chpt}
\end{equation}
Here ${\cal V}$ describes the lattice artifacts. For staggered fermions
these are dominated by taste breaking terms \cite{SCHiPT}
\begin{equation}
{\cal V} = - \frac{1}{2} C_4 \Tr \left( \xi_{\mu 5} U \xi_{5 \mu}
 U^\dagger + h.c. \right) + \dots ~,
\label{eq:C4_chpt}
\end{equation}
where we displayed only the term that dominates the pseudoscalar mass
splittings explicitly. Here $\xi_\mu = \gamma_\mu^\ast$ are taste matrices.
This is also the term that dominates the lattice effects in the low-lying
Dirac spectrum in the regime of weak taste breaking.

For Wilson fermions, the lattice artifact terms are \cite{WCHiPT}
\begin{equation}
{\cal V} = W_8 \Tr \left( U^2 + U^{\dagger 2} \right)
 + W_6 \left[ \Tr \left( U + U^\dagger \right) \right]^2
 + W_7 \left[ \Tr \left( U - U^\dagger \right) \right]^2 ~.
\label{eq:W_chpt}
\end{equation}
The two-trace terms, with coefficients $W_6$ and $W_7$, are suppressed at
large $N_c$. We will neglect them here and only keep the one-trace
correction term, proportional to $W_8$.

\section{Random matrix theory including lattice artifacts}

The $\epsilon$-regime of \chpt\, at leading order, can equivalently be
described by a chiral random matrix theory (RMT). For continuum QCD, the
Dirac operator is represented in RMT as
\begin{equation}
{\cal D}_0 = \left( \begin{array}{cc}
0 & i W \\ i W^\dagger & 0
\end{array} \right)
\label{eq:D_RMT}
\end{equation}
with $W$ a random $(N+\nu) \times N$ complex matrix, when working in
a sector with index (topological charge) $\nu$.

\subsection{Staggered RMT}

For staggered fermions, the Dirac operator in staggered RMT (SRMT) is
represented as \cite{SRMT}
\begin{equation}
{\cal D}_{stag} = {\cal D}_0 \otimes \mathbb{I}_4 + a {\cal T} ~,
\qquad 
{\cal T}_{C_4} = \left( \begin{array}{cc}
A_\mu & 0 \\ 0 & B_\mu
\end{array} \right) \otimes \xi_{\mu 5} ~.
\label{eq:D_SRMT}
\end{equation}
Here $\mathbb{I}_4$ is the identity matrix in taste space and
${\cal T}$ denotes the taste beaking terms.
The dominant taste-breaking term, corresponding to the term explicitly
shown in Eq.~(\ref{eq:C4_chpt}), denoted by ${\cal T}_{C_4}$ is shown
in the second part. There, $A_\mu$ and $B_\mu$ are random Hermitian
matrices of size $(N+\nu) \times (N+\nu)$ and $N \times N$, respectively.
The width of the Gaussion distributions of $A_\mu$ and $B_\mu$ is
proportinal to $C_4$. The dimensionless combination $a^2 C_4 V$ controls
the strength of the taste breaking in SRMT.

Without the taste-breaking terms $a {\cal T}$ ({\it i.e.,} in the
continuum limit) the eigenvalues of the Dirac operator come in
degenerate quartets. For weak taste breaking, $a^2 C_4 V \ll 1$, the
quartets of eigenvalues are split, at leading order in a perturbation
by $a {\cal T}_{C_4}$, into pairs of eigenvalues with splitting
$\Delta \lambda_{quart}$ \cite{SRMT}.
The pairs of eigenvalues, in turn, are split, for weak taste breaking,
at second order in a perturbation by $a {\cal T}_{C_4}$, with subdominant
splitting $\Delta \lambda_{pair}$. So SRMT predicts
\begin{equation}
\frac{\Delta \lambda_{quart}}{\lambda} \propto a \sqrt{C_4 V} ~,
\qquad
\frac{\Delta \lambda_{pair}}{\lambda} \propto a^2 C_4 V ~.
\label{eq:split_pair}
\end{equation}

\subsection{Wilson RMT}

For Wilson fermions, the Dirac operator in Wilson RMT (WRMT) is
represented as \cite{WRMT}
\begin{equation}
{\cal D}_W = {\cal D}_0 + \tilde{a} \left( \begin{array}{cc}
A & 0 \\ 0 & B \end{array} \right) ~,
\label{eq:D_WRMT}
\end{equation}
with $A$ and $B$ random Hermitian matrices of size $(N+\nu) \times (N+\nu)$
and $N \times N$, respectively, that represent the chiral symmetry breaking
Wilson term of the lattice Wilson-Dirac operator.

Akemann {\it et al.}, Ref.~\cite{WRMT}, have worked out the
eigenvalue distribution of the Hermitian Wilson-Dirac operator,
$H_W = \gamma_5 \left( D_W + m_0 \right)$, or its RMT equivalent,
${\cal H}_W = \gamma_5 \left( {\cal D}_W + \tilde{m} \right)$ with
\begin{equation}
\hat{m} = m \Sigma V = 2 \tilde{m} N \quad \text{and} \quad
\hat{a}^2 = a^2 W_8 V = \frac{1}{2} \tilde{a}^2 N
\label{eq:WRMT_params}
\end{equation}
held fixed, using Wilson \chpt. The results were reproduced directly from
WRMT in \cite{WRMT2}. In Eq.~(\ref{eq:WRMT_params}) $m$ is a suitably
subtracted version of the bare mass $m_0$. In the analytical predictions,
the eigenvalues are rescaled with $\Sigma V$ in the lattice QCD case, and
with $2N$ for WRMT.

\section{Results for staggered fermions compared to SRMT}

\begin{figure}[h]
\begin{center}
\begin{tabular}{c c}
\includegraphics[width=7.0cm]{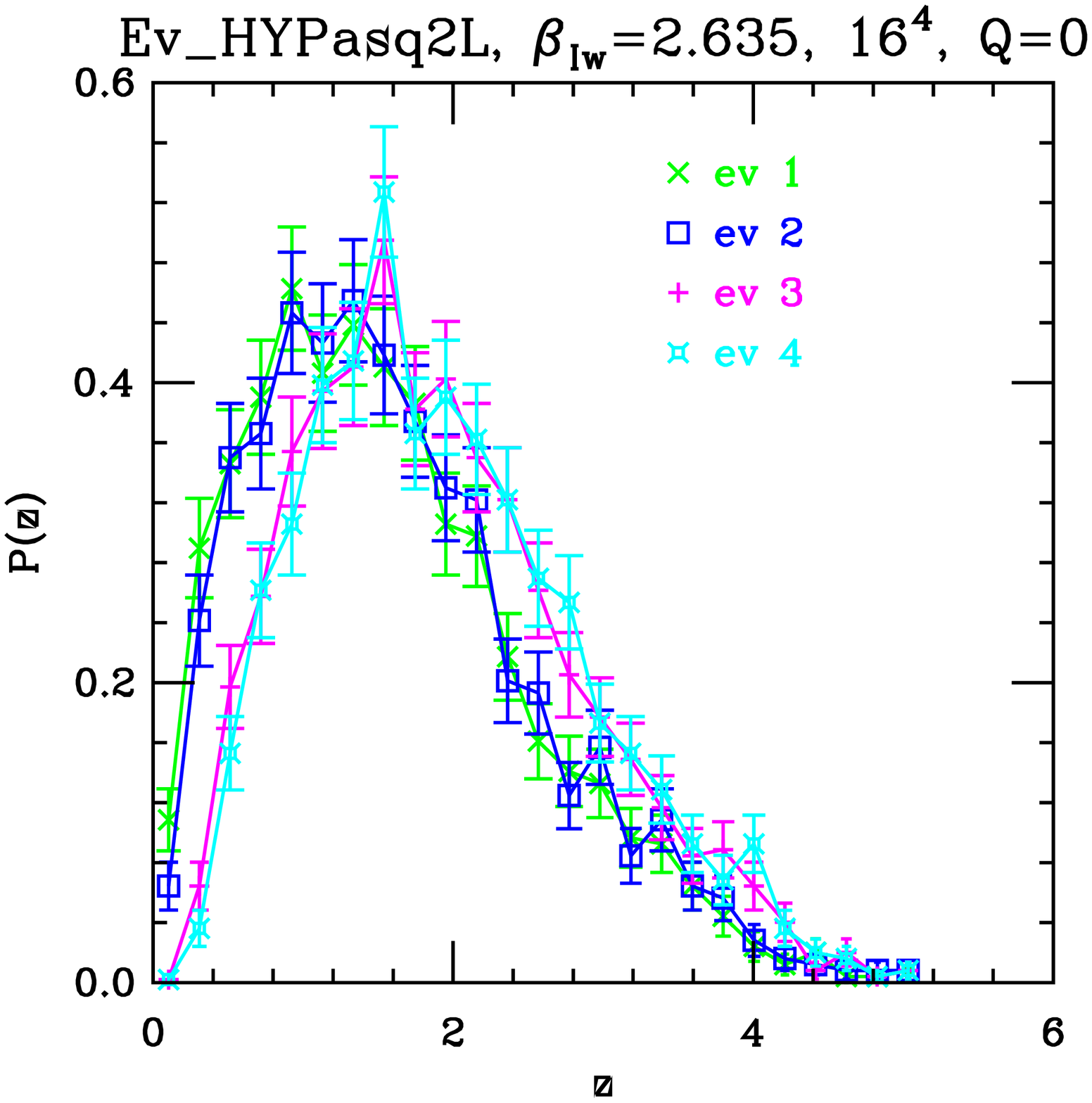}
&
\includegraphics[width=7.0cm]{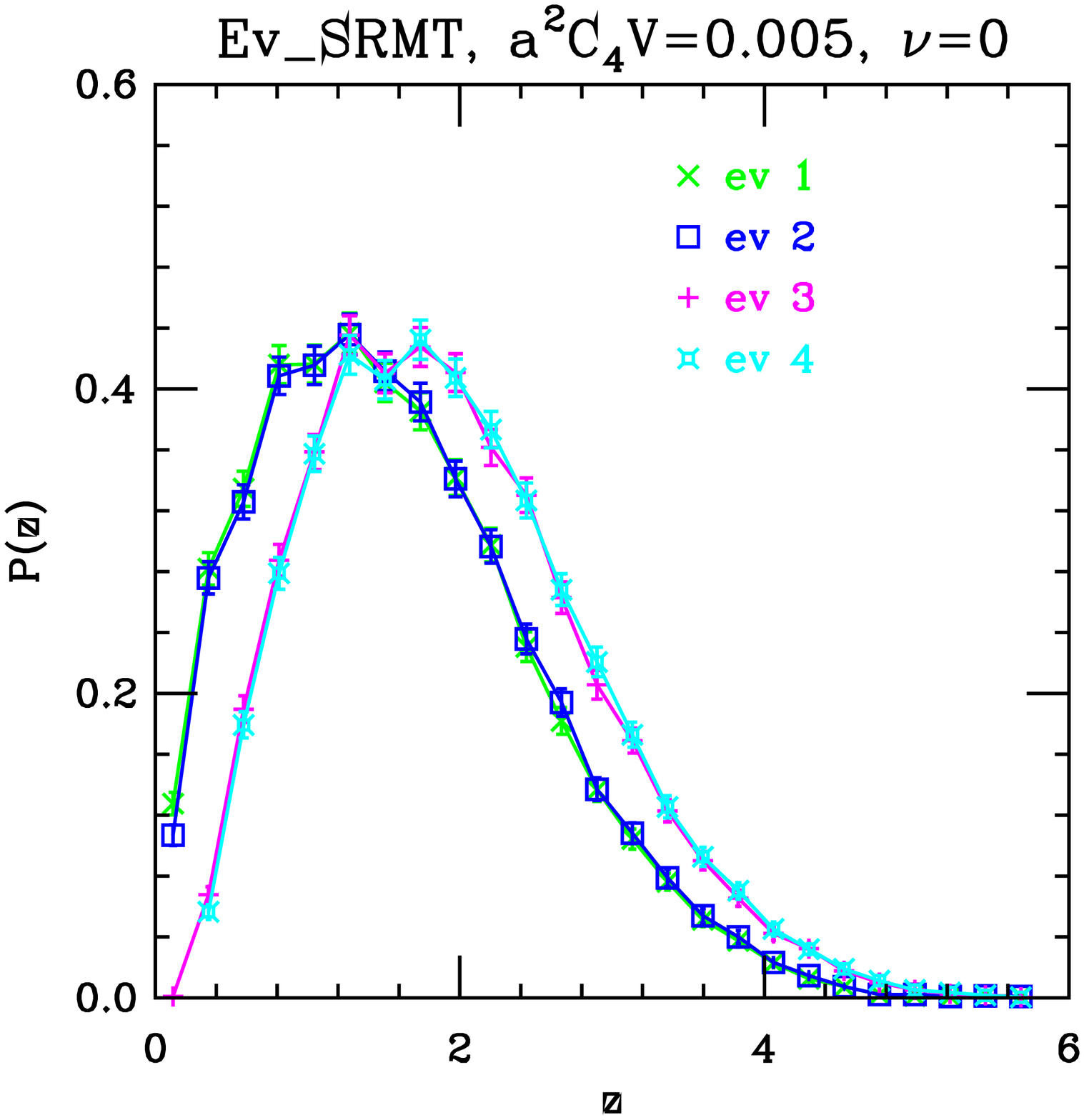}
\end{tabular}
\end{center}
\vspace{-0.6truecm}
\caption{Distributions of the lowest four (rescaled) eigenvalues
for $Q=0$.  The left panel shows the staggered Dirac eigenvalues
from an ensemble of $a=0.093$ fm and $L=1.5$ fm configurations, the right
panel the eigenvalues from a MC generated SRMT ensemble with taste-breaking
parameter $a^2 C_4 V = 0.005$.}
\label{fig:Q0_stag_dist}
\end{figure}

We have computed low lying staggered eigenvalues for various staggered
actions with different smearings using ensembles of pure gauge
configurations generated with the Iwasaki gauge action. As observed
previously \cite{StagEvs}, the quartet structure, and the appearance of
clearly distinct would-be zeromodes, becomes more visible with increased
smearing and with smaller lattice spacing. Here we show and discuss some
results with our best staggered action, a HISQ-like action but with first a
HYP(ii) smearing step \cite{HYPii} instead of a fat7 smearing step and
subsequent unitarization, followed by an asq step with the appropriate
Lepage term, denoted by ``HYPasq2L''. For more details and further results
we refer to Ref.~\cite{Urs_StagEig}. Here we illustrate a few of our
findings.

We compare distributions of the lowest four eigenvalues from $\sim 1200$
$Q=0$ configurations at $a=0.093$ fm and $L=1.5$ fm with Monte Carlo
generated SRMT $\nu=0$ eigenvalue distributions for the taste violating
parameter $a^2 C_4 V = 0.005$ in Fig.~\ref{fig:Q0_stag_dist}.\footnote{We
thank James Osborn for providing the eigenvalues from SRMT generated by
MC.} The LQCD eigenvalues are rescaled by $\Sigma V$ and the SRMT ones by
$2N$ for the comparison.  The distributions agree quite nicely.

\begin{figure}[h]
\begin{center}
\begin{tabular}{c c}
\includegraphics[width=7.0cm]{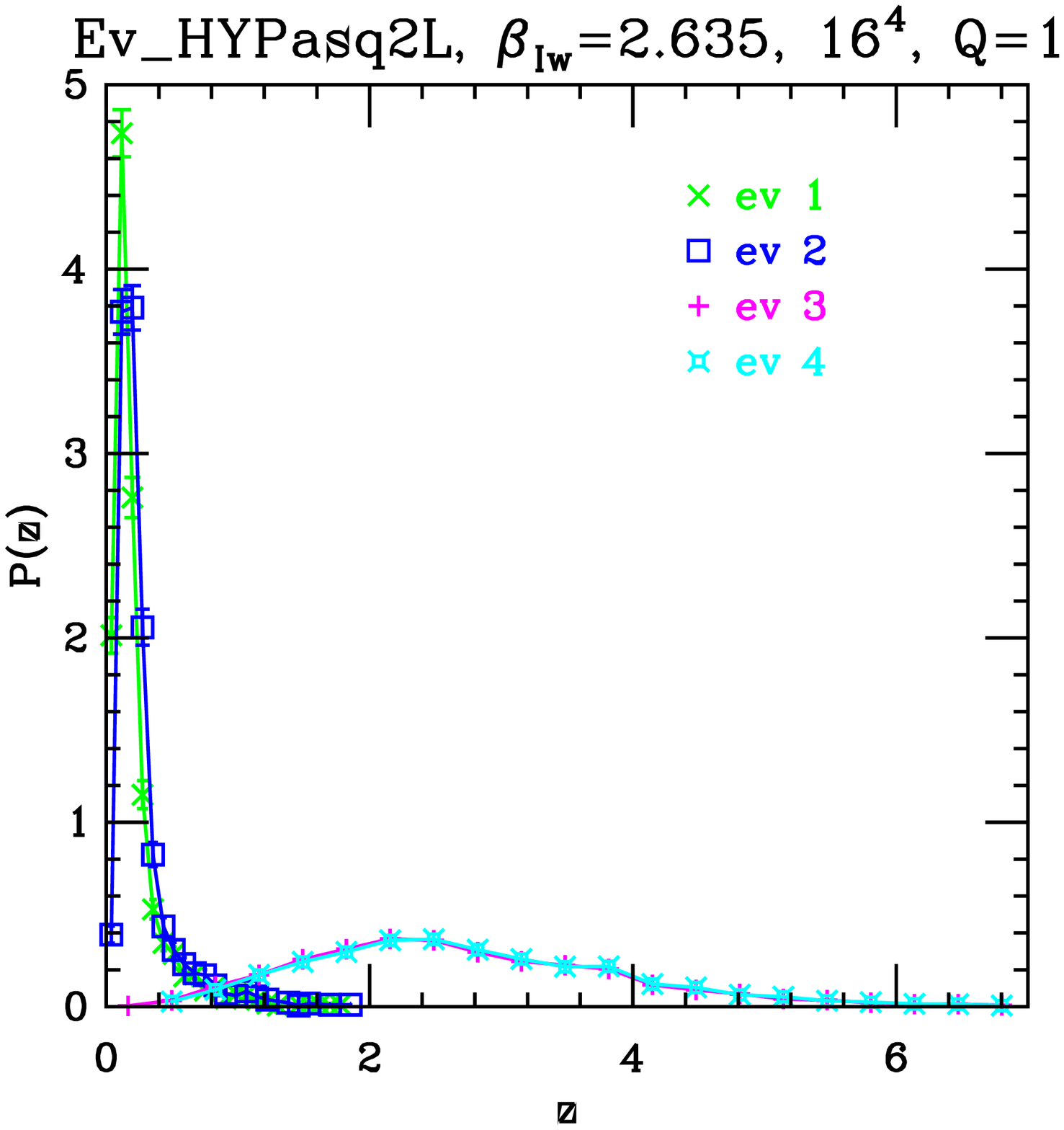}
&
\includegraphics[width=7.0cm]{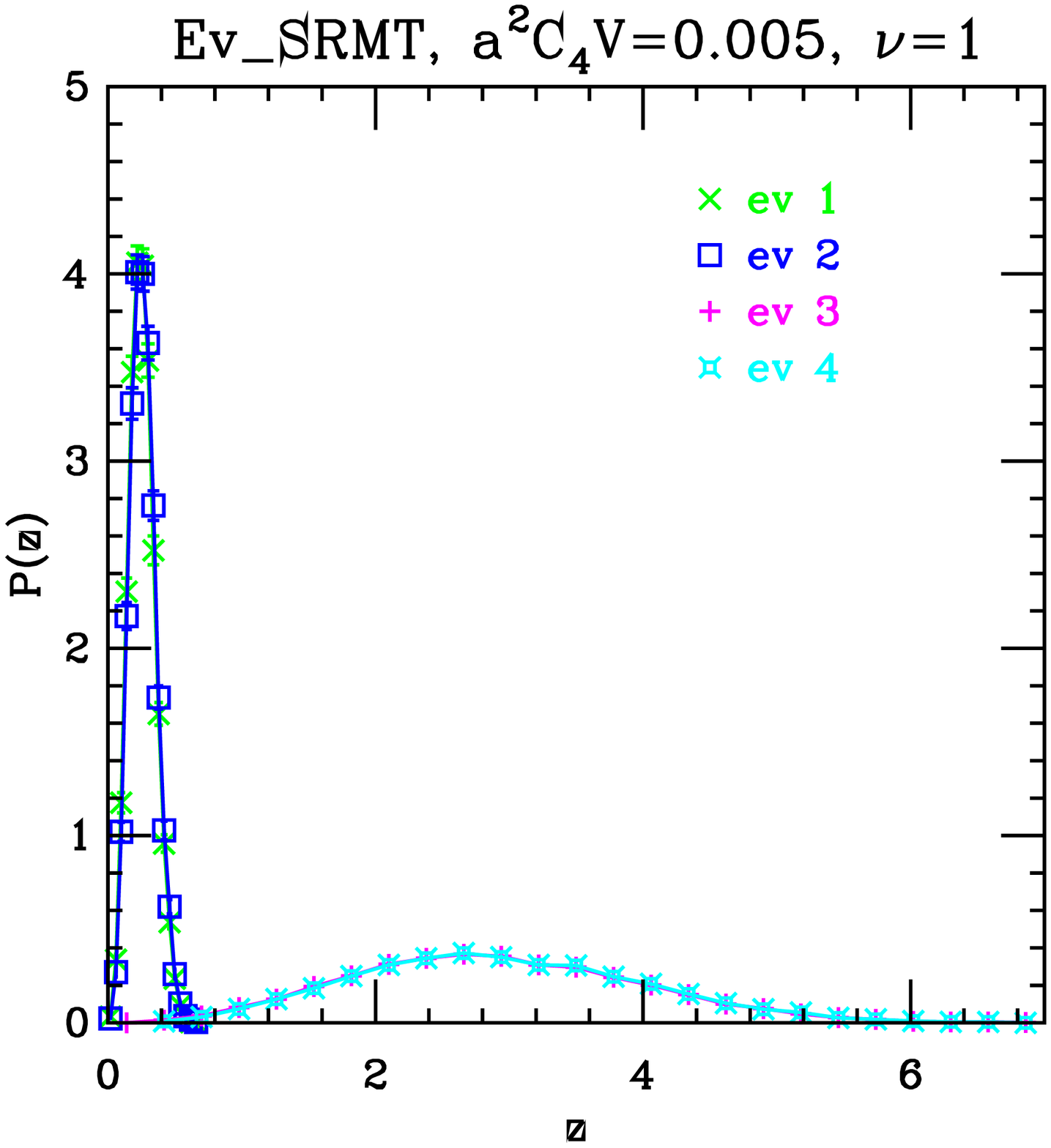}
\end{tabular}
\end{center}
\vspace{-0.6truecm}
\caption{Same as Fig.~\protect\ref{fig:Q0_stag_dist} but for $|Q|=1$
($\nu=1$) configurations. Here, the lowest two eigenvalues are the
would-be zeromodes.}
\label{fig:Q1_stag_dist}
\end{figure}

Fig.~\ref{fig:Q1_stag_dist} shows the distribution of the lowest four
eigenvalues, including the two positive would-be zeromodes, for $\sim 2000$
$|Q|=1$ configurations of the same gauge ensemble and $\nu=1$ SRMT MC
generated eigenvalue distributions. Again, the distributions agree quite
nicely, although the distributions of the would-be zeromodes for the
staggered Dirac operator have somewhat longer tails than their SRMT
counterparts.  In the SRMT distributions, the peak height of the ``would-be
zeromodes'' turns out to be the feature most sensitive to the value of the
taste-breaking parameter $a^2 C_4 V$. It is from this peak height that we
estimated that $a^2 C_4 V = 0.005$ provides the best match to the LQCD data.
 
\begin{figure}[h]
\begin{center}
\begin{tabular}{c c}
\includegraphics[width=7.0cm]{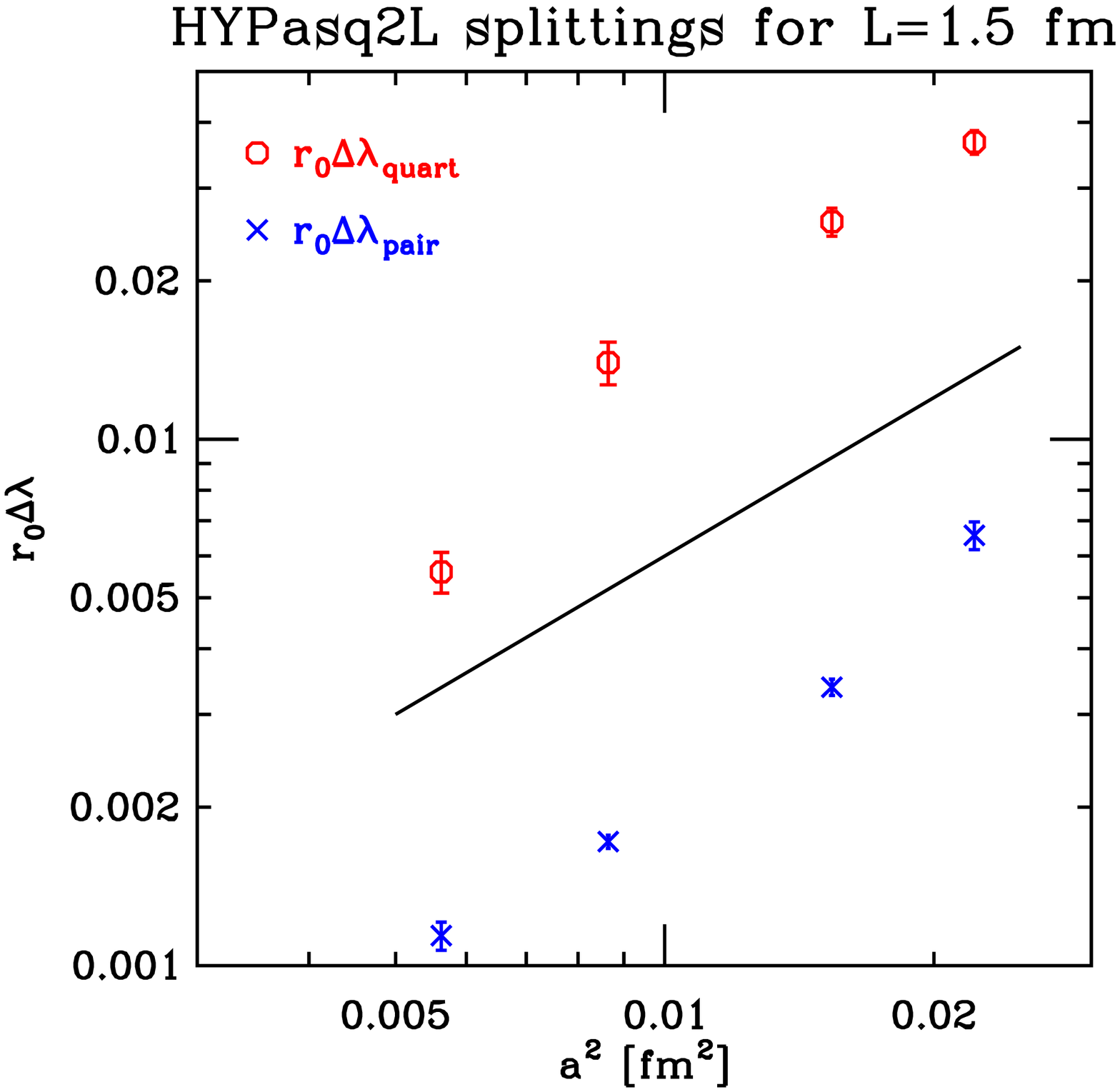}
&
\includegraphics[width=7.0cm]{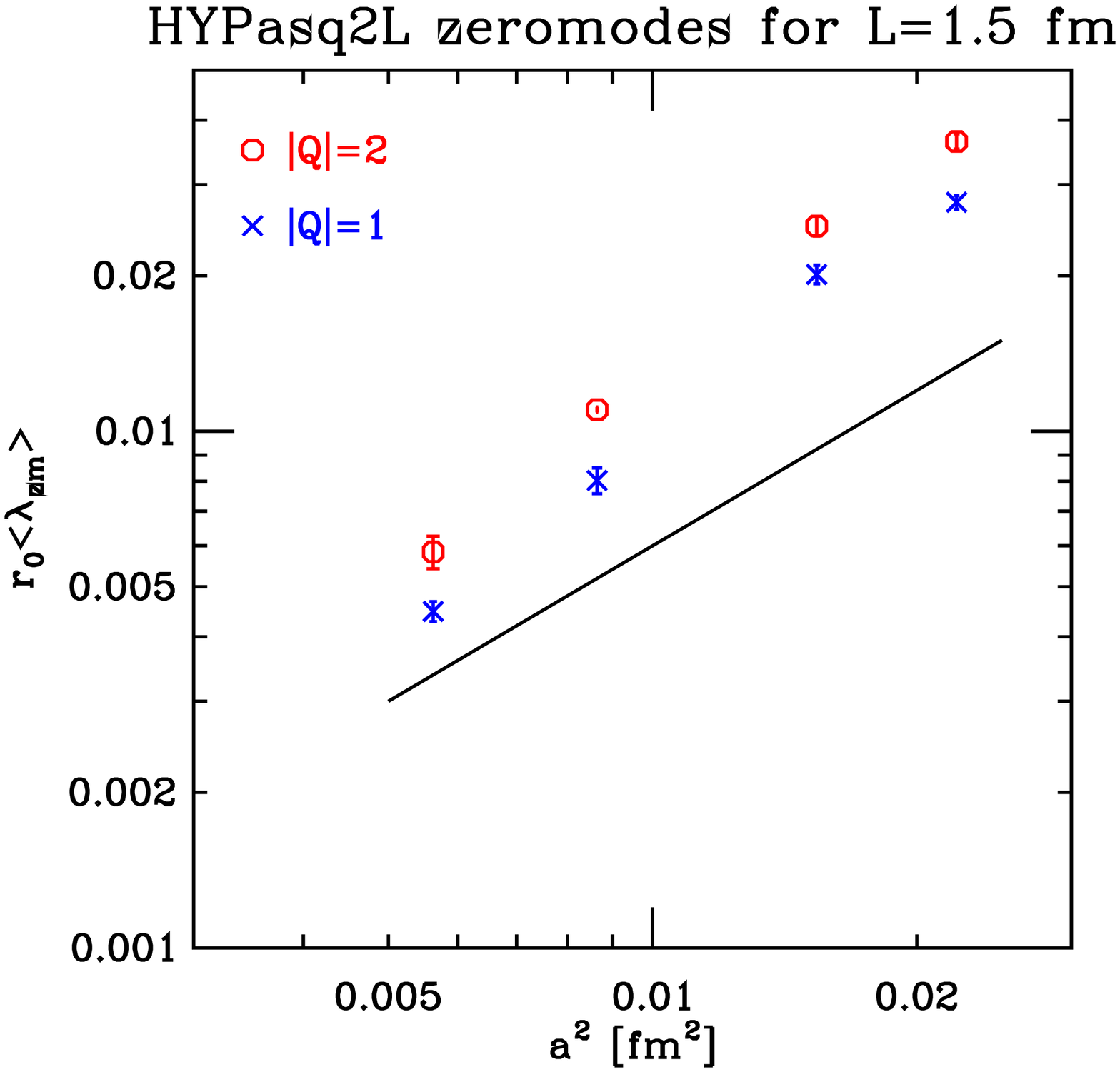}
\\
\includegraphics[width=7.0cm]{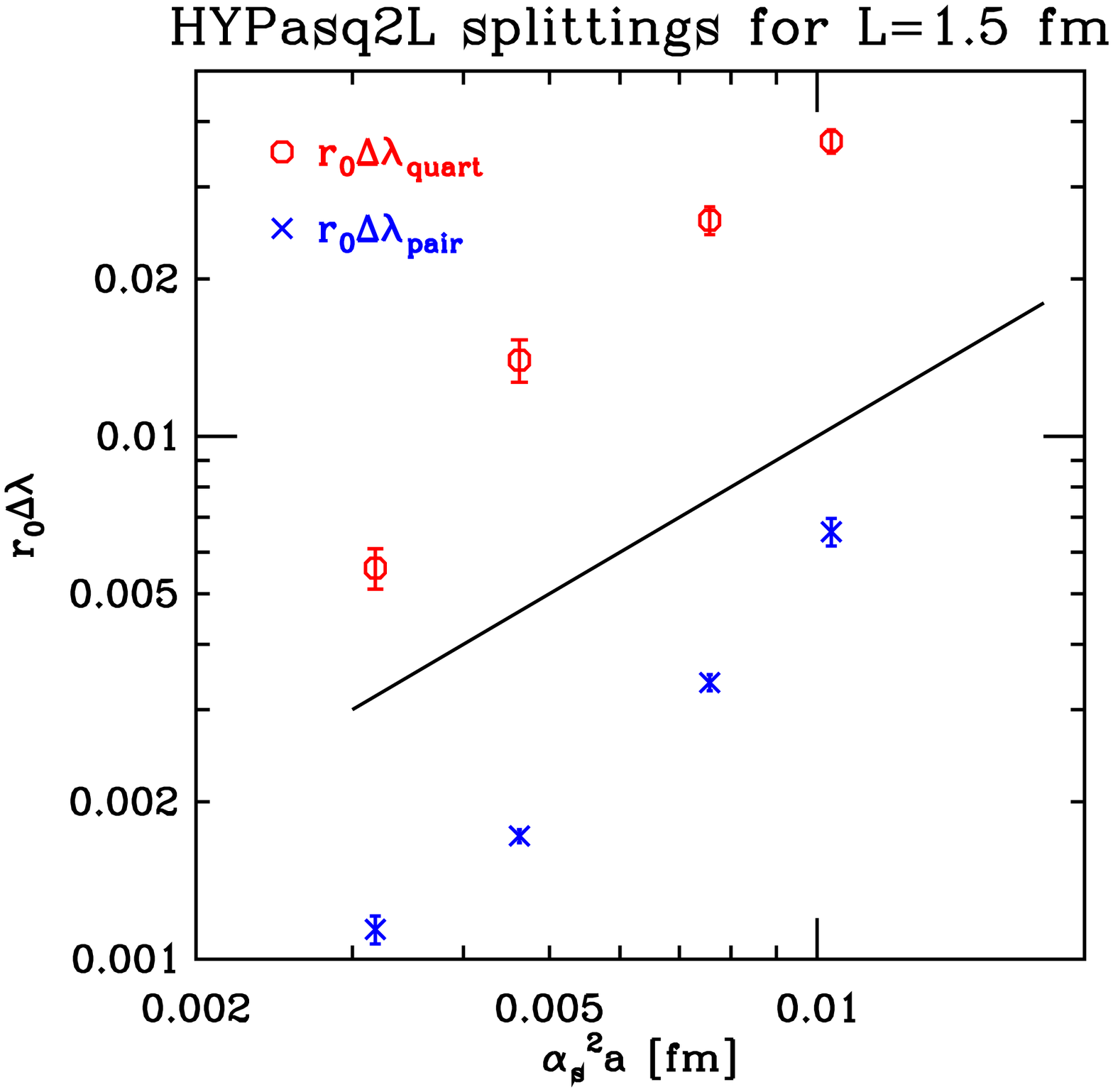}
&
\includegraphics[width=7.0cm]{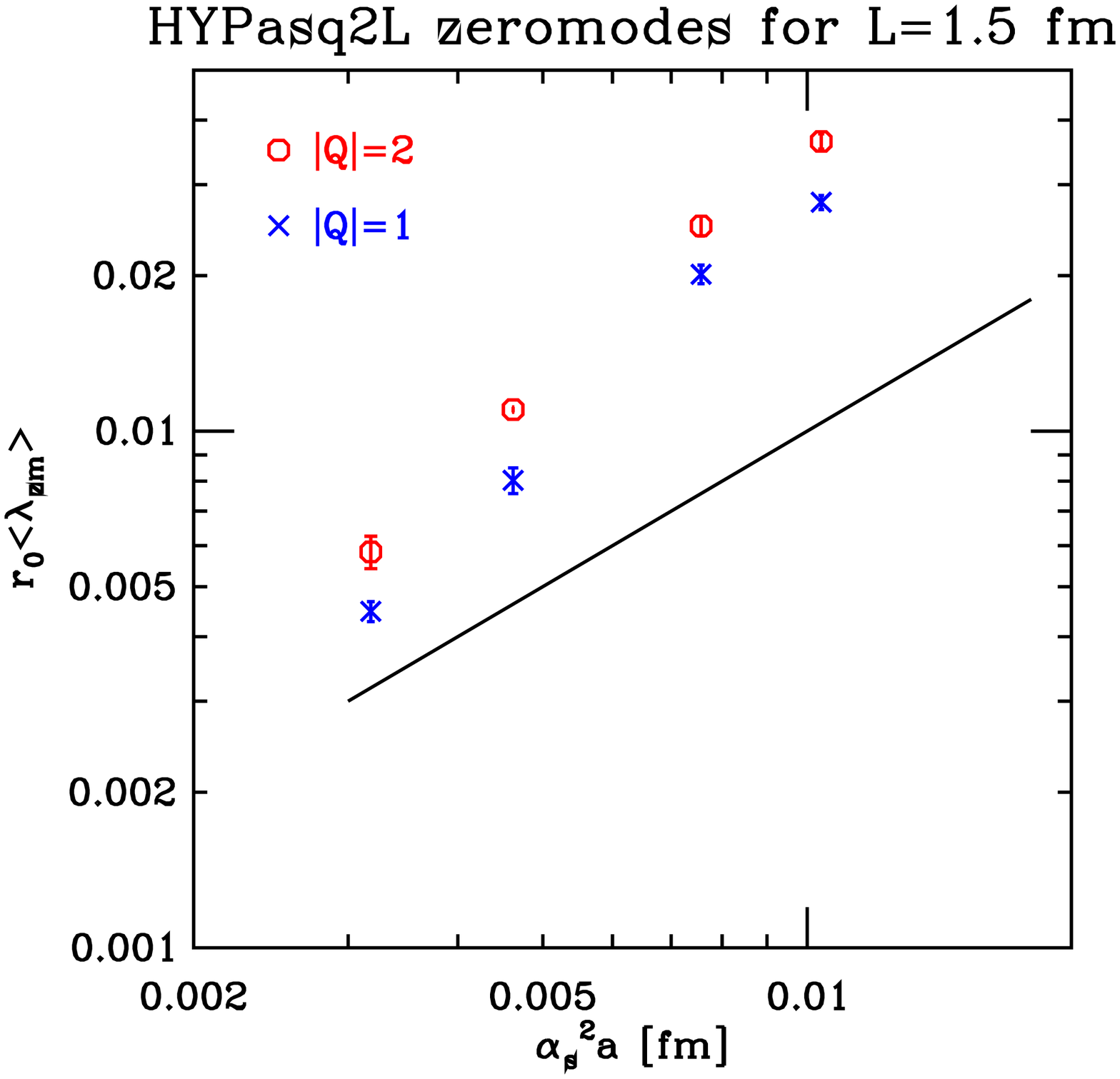}
\end{tabular}
\end{center}
\vspace{-0.6truecm}
\caption{The vanishing of the splitting (left panels) and would-be
zeromodes (right panels) -- shown is the average over the would-be
zeromodes with slightly positive eigenvalues -- as a function of
$a^2$ (top) or $\alpha_s^2 a$ (bottom). The inclined straight lines
show the behavior to be tested in the plots.}
\label{fig:stag_scaling}
\end{figure}

According to the SRMT prediction, Eq.~(\ref{eq:split_pair}), the dominant
splitting $\Delta \lambda_{quart}$ should vanish, at fixed volume, like
$\alpha_s^n a$, where $n$ depends on the degree of improvement, while the
subdominat splitting $\Delta \lambda_{pair}$ should vanish like $a^2$. We
compare these predictions, as well as the vanishing of the would-be
zeromodes, for fixed lattice size $L=1.5$ fm, in Fig.~\ref{fig:stag_scaling}.
The would-be zeromodes appear to vanish a little faster than $a^2$.  The
behavior of the splittings is less clear, with a vanishing as $a^2$ favored
for the smaller splitting, $\Delta \lambda_{pair}$. But both splittings
certainly vanish in the continuum limit, as expected.

\section{Results for Wilson fermions compared to WRMT\protect\footnote{The
results in this section were obtained in collaboration with Poul Damgaard
and Kim Splittorff.}}

\begin{figure}[h]
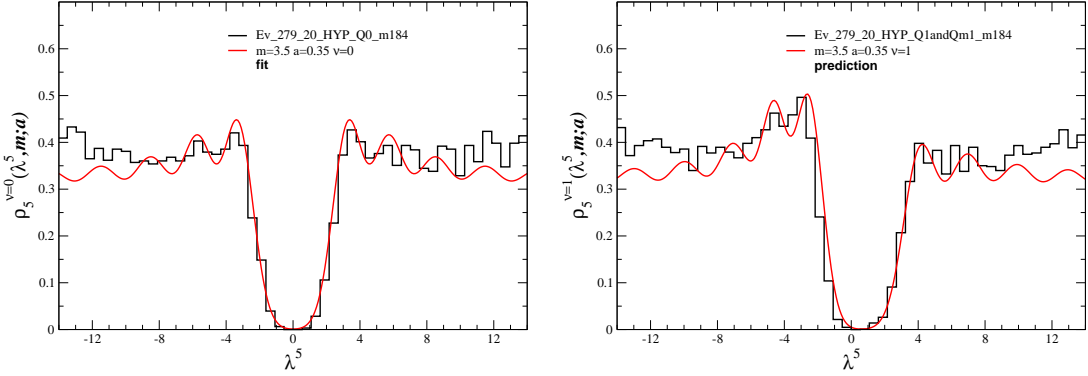

\begin{center}
\begin{tabular}{c c}
\includegraphics[width=7.0cm]{Ev_279_20_HYP_Q0_m184-1172configs-fit2-v2.eps}
&
\includegraphics[width=7.0cm]{Ev_279_20_HYP_Q1andQm1_m184-990resp988configs-fit2-v2.eps}
\end{tabular}
\end{center}
\vspace{-0.6truecm}
\caption{The eigenvalues density of $H_w$ with bare mass $am_0=-0.184$ on
an ensemble of $a = 0.075$ fm and $L = 1.5$ fm configurations with
$Q=0$ (left) and $|Q|=1$ (right). The $Q=0$ distribution was used to
obtain the WRMT parameters. The same parameters are used for the $|Q|=1$
WRMT distribution on the right.}
\label{fig:Ev_Wils_m184}
\end{figure}

We computed the low-lying eigenvalues of the Hermitian Wilson-Dirac
operator $H_W$ on the $Q=0$ and $|Q|=1$ configurations of an ensemble with
lattice spacing $a= 0.075 $ fm and size $L=1.5$ fm, as before generated
with the Iwasaki gauge action, for two bare quark masses. To suppress
dislocations we smoothened the gauge field with one HYP smearing step
before constructing the Wilson-Dirac operator.

RMT predictions are made for sectors with a given index $\nu$. Such an
index can be defined from the real eigenvalues of the Wilson-Dirac operator
and the chiralities of the corresponding eigenmodes Ref.~\cite{WRMT}.
Equivalently, the index can be obtained from the spectral flow of the
Hermitian Wilson-Dirac operator \cite{spec_flow}. In both cases the
definition depends on a cut: the maximum real eigenvalue kept or the mass
at which the flow is terminated. For our setup we found little dependence
on this cut and good agreement with the cheaper definition of topological
charge with six HYP smearings and use of an improved $F {}^*F$ operator
\cite{Boulder_Q}.

To compare the sepctrum of the Hermitian Wilson-Dirac operator with
predictions from WRMT, we used the distribution in the $Q=0$ sector with
the bare mass corresponding to a lighter quark mass to find the WRMT
parameters $\hat{m}$ and $\hat{a}$, and the eigenvalue rescaling factor
$\Sigma V$ that best describe our data, see Fig.~\ref{fig:Ev_Wils_m184}
(left).  Using the {\it same} parameter values, WRMT then predicts the
$|Q|=1$ distribution that can be compared with the measured one, see
Fig.~\ref{fig:Ev_Wils_m184} (right). We find a nice agreement with the
measured (histogrammed) distribution.

\begin{figure}[h]
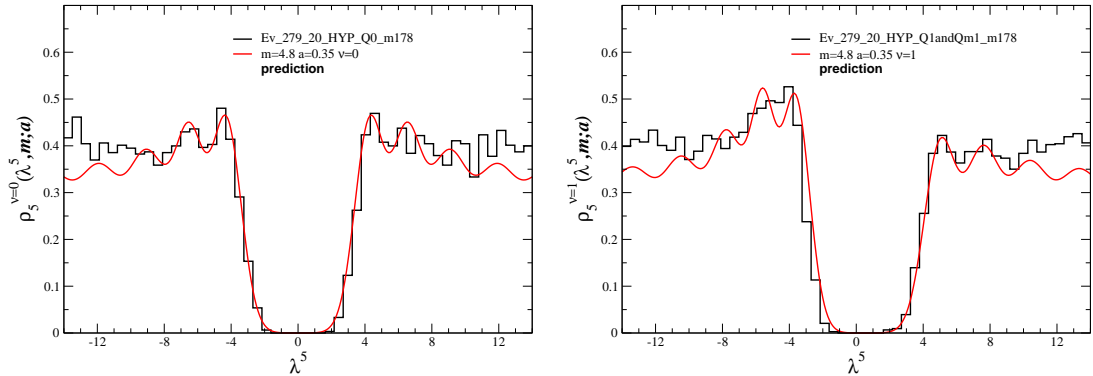

\vspace{0.6truecm}
\begin{center}
\begin{tabular}{c c}
\includegraphics[width=7.0cm]{Ev_279_20_HYP_Q0_m178-1172configs-m-scaling-v1.eps}
&
\includegraphics[width=7.0cm]{Ev_279_20_HYP_Q1andQm1_m178-990resp988configs-fit2-m-scaling.eps}
\end{tabular}
\end{center}
\vspace{-0.6truecm}
\caption{Same as Fig.~\protect\ref{fig:Ev_Wils_m184} but for bare mass
$am_0 = -0.178$. The same WRMT parameters, up to changing $\hat{m}$
according to the bare mass difference, $\delta\hat{m} = \delta m_0
\Sigma V$, were used.}
\label{fig:Ev_Wils_m178}
\end{figure}

Changing only the bare quark mass in $H_W$ should leave the WRMT parameter
$\hat{a}$, and the rescaling factor $\Sigma V$ unchanged, while the WRMT
parameter $\hat{m}$ should be changed by $\delta \hat{m} = \delta m_0
\Sigma V$, Eq.~(\ref{eq:WRMT_params}), where $\delta m_0$ is the difference
in bare quark mass. We, therefore, have parameter free predicitions for the
eigenvalue distributions with the second quark mass. As shown in
Fig.~\ref{fig:Ev_Wils_m178}, these predictions work well.

\begin{figure}[h]
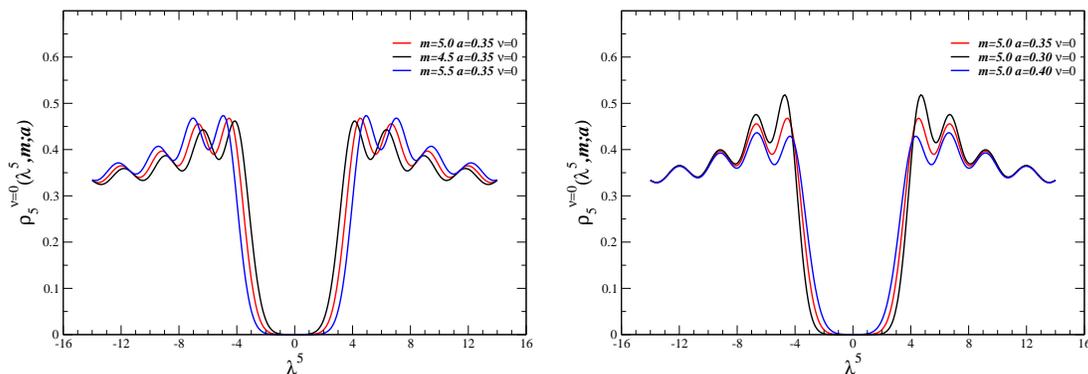

\vspace{0.6truecm}
\begin{center}
\begin{tabular}{c c}
\includegraphics[width=7.0cm]{mdep.eps}
&
\includegraphics[width=7.0cm]{adep.eps}
\end{tabular}
\end{center}
\vspace{-0.8truecm}
\caption{The dependence of the distribuion of the eigenvalue density of
$\cal{H}_W$ on $\hat{m}$ (left) and on $\hat{a}$ (right).}
\label{fig:WRMT_dependence}
\end{figure}

We illustrate the sensitivity of the analytic WRMT distributions to the
WRMT parameters $\hat{m}$ in Fig.~\ref{fig:WRMT_dependence} (left) and to
$\hat{a}$ in Fig.~\ref{fig:WRMT_dependence} (right). For the ensemble
considered in Figs.~\ref{fig:Ev_Wils_m184} and \ref{fig:Ev_Wils_m178} we
can obtain $\hat{m}$ to an accuracy of about $0.5$ and $\hat{a}$ to about
$0.05$. Going to smaller lattice spacing will increase the accuracy.  For
further details and more tests, see \cite{WRMT_num}.  A.~Deuzeman presented
results from a similar study at this conference with compatible results
\cite{WRMT_num2}.

\end{document}